\documentclass[12pt,preprint]{aastex}

\begin{document}



\title {The Structure of the X-ray and Optical Emitting Regions of the Lensed Quasar Q~2237+0305}


\author{A. M. Mosquera$^{1}$, C. S. Kochanek$^{1,2}$, B. Chen$^{3}$, X. Dai$^{3}$, J. A. Blackburne$^{1}$, G. Chartas$^{4}$}

\bigskip

\affil{$^{1}$Department of Astronomy, The Ohio State University, 140 West 18th Avenue, Columbus, 
OH 43210, USA}
\affil{$^{2}$Center for Cosmology and Astroparticle Physics, The Ohio State University,
191 West Woodruff Avenue, Columbus, OH 43210, USA}
\affil{$^{3}$Homer L. Dodge Department of Physics and Astronomy, The University of Oklahoma, Norman, OK, 73019, USA}
\affil{$^{4}$Department of Physics and Astronomy, College of Charleston, Charleston, SC, 29424, USA}


\begin{abstract}
 
We use gravitational microlensing to determine the size of the X-ray and optical 
emission regions of the quadruple lens system Q~2237+0305. The optical half-light radius, 
$\mathrm{log}(R_{1/2,V}/\mathrm{cm})=16.41\pm0.18$ (at $\lambda_{rest}=2018 $\AA), is significantly larger 
than the observed soft, $\mathrm{log}(R_{1/2,soft}/\mathrm{cm})=15.76^{+0.41}_{-0.34}$ 
($1.1-3.5$ keV in the rest frame), and hard, $\mathrm{log}(R_{1/2,hard}/\mathrm{cm})=15.46^{+0.34}_{-0.29}$ 
($3.5-21.5$ keV in the rest frame), band X-ray emission.
There is a weak evidence that the hard component is more compact than the 
soft, with $\mathrm{log}(R_{1/2,soft}/R_{1/2,hard})\simeq 0.30^{+0.53}_{-0.45}$. 
This wavelength-dependent structure agrees with recent results found in other lens systems using 
microlensing techniques, and favors geometries in which the corona is concentrated near the inner edge of the 
accretion disk. While the available measurements are limited, the size of the X-ray emission region appears to be 
roughly proportional to the mass of the central black hole.

\end{abstract}



\section{Introduction}

AGNs play a crucial role in the evolution of the Universe, but their internal structure remains 
poorly understood. The energy is generated by accretion onto a supermassive black hole (SMBH) 
through a disk, but the detailed structure of the disk \citep[e.g.,][]{blaes04, blaes07}, the origin of 
the non-thermal emissions \citep[]{reynolds}, and the source of any outflows \citep{Martin2000} are all topics of debate. 
Although simple thin accretion disk models describe some of the observed behavior, they do not reproduce all the observed 
features \citep[e.g.,][and references therein]{blaes07}. Testing accretion disk models is challenging in large 
part because of our inability to resolve the emission regions. 
Reverberation mapping techniques \citep[e.g.,][]{Brad93} strongly constrain the structure of the distant emission line 
regions, but not the continuum emission regions beyond the limits already set by time 
variability \citep[e.g.,][for modern models of optical variability]{Kelly2009, Szy2010, macleod2010}.
Fortunately, gravitationally lensed quasars provide us with a unique tool to zoom in on 
the structure of AGN and explore their physics in more detail - a ``telescope'' that only becomes more powerful as the 
target source becomes smaller.

Microlensing magnification is caused by stars and white dwarfs in the lens galaxy close to the line of 
sight towards the lensed quasar images. It leads to uncorrelated flux variations due to the relative motions 
of the quasar, the lens, its stars, and the observer \citep[see the review by][]{joachim-saas06}. The 
microlensing signal depends on the structural and dynamical properties of the source and the lens. Since the 
magnification depends upon the size of the source, the simplest application of quasar microlensing is to measure 
the size of quasar accretion disks. In essence, the 
amplitude of the microlensing variability encodes the disk size, with larger disks showing lower variability
amplitudes. Given adequate light curves, this is now routine, with studies of the
scaling of size with black hole mass \citep{morgan10}, 
wavelength \citep{timo2008,bate08,eigenbrod08,poindexter08,Floyd2009,ana09,ana11a,evencio11,jeff2011,Motta2012}
and the nature of the emitting source \citep[thermal or non-thermal X-ray emission,][]{pooley07,morgan08,
chartas09,dai10,jeff2011,jeff2012f,jeff1104,morgan12}. Microlensing has also been used to explore the 
spatial structure of the broad line region (BLR) of quasars 
\citep[e.g.,][]{Lewis98,Cristina02,Sluse07,Sluse11,ODowd10,guerras2012}, although the amplitudes of BLR lensing effects 
will be small due to its large size \citep[see][]{ana11b}.

The largest microlensing variability amplitudes are observed in the X-rays, as emphasized by \cite{chartas02}, 
\cite{dai2003}, and \cite{pooley07}, indicating that this emission must arise close to the inner edge of the disk. Quantitative 
microlensing X-ray size estimates are now available for several systems \citep[][]{dai10, morgan12, jeff2012f, jeff1104}, although the 
sparse data means that the measurements are largely upper limits because the likelihood distributions for the lower limits 
are prior dependent. 

The well-known quadruple lensed quasar Q~2237$+$0305 \citep{Huchra85} remains the best system for studying microlensing phenomena.  
In this system a $z_{S}=1.69$ background quasar is lensed by a nearby spiral galaxy at $z_L=0.039$ 
to form 4 images of the quasar. The images lie in the bulge of the galaxy, where the optical depth for microlensing is 
high and there is little dark matter. Furthermore, the low lens redshift leads to unusually short microlensing variability time scales ($\sim$ months). 
Microlensing fluctuations were first detected by \cite{irwin89} and have now been monitored for well over a decade, 
principally by the Optical Gravitational Lensing Experiment(OGLE) \citep{wozniak2000, Udalski}.
Moreover, due to the symmetry of the system, and the proximity of the lens galaxy, time delay corrections 
are negligible \citep[$<1$ day,][]{dai2003,Vakulik,koptelova06} which simplifies analysis. Because of these characteristics, 
Q~2237$+$0305 has been the object of many studies, and monitored in different 
bands \citep[e.g.,][]{wozniak2000,dai2003,Vakulik, timo2008,eigenbrod08, ana09,ODowd10}. Our X-ray studies 
\citep{chen2011, chen2012}, some of which were independently analyzed by \cite{Zimmer2011}, not only found evidence for 
microlensing variability but also for a dependence upon the X-ray continuum energy.

In this paper we present a complete analysis of the OGLE and \cite{chen2011, chen2012} {\it Chandra} data to set constraints 
on the structure of the optical and X-ray emitting regions. We also compare our X-ray size estimates to those obtained 
for HE 0435$-$1223 \citep{jeff2012f},  QJ 0158-4325 \citep{morgan12}, RXJ 1131$-$1231 \citep{dai10} and HE 1104$-$1805 (Blackburne et al. 2012), 
to explore possible correlations of the X-ray sizes with BH mass. A summary of the observations, a description of the lightcurve fitting method, 
and a discussion of the main results is presented in Section \ref{sec2}. In Section \ref{sec3} we summarize the results and their 
implications. Throughout this work we assume $\Omega_m=0.3$, $\Omega_{\Lambda}=0.7$, and $H_0=72$ km sec$^{-1}$ Mpc$^{-1}$.

\section{Data Analysis and Discussion}\label{sec2}

\subsection{Observations}\label{ssec2}

Q~2237$+$0305 was observed in X-rays with ACIS \citep{garmire2003}
on the {\it Chandra X-ray Observatory} \citep[{\it Chandra},][]{chandra}. Combined with archival data, 
we have a total of 20 epochs between JD 2251 and JD 5529 taken during Cycles 1, 2, 7 and 11. 
The full band was defined as 0.4$-$8.0 keV (in the observers frame). These were then divided into 
soft (0.4$-$1.3 kev) and hard (1.3$-$8 keV) bands, where the division at 1.3 keV was chosen to balance 
the counts between the bands. A detailed description of the X-ray observations and data reduction techniques 
can be found in \cite{chen2011, chen2012}.

\begin{figure}
\begin{center}
\vspace{0.5 cm}
\includegraphics[scale=0.8]{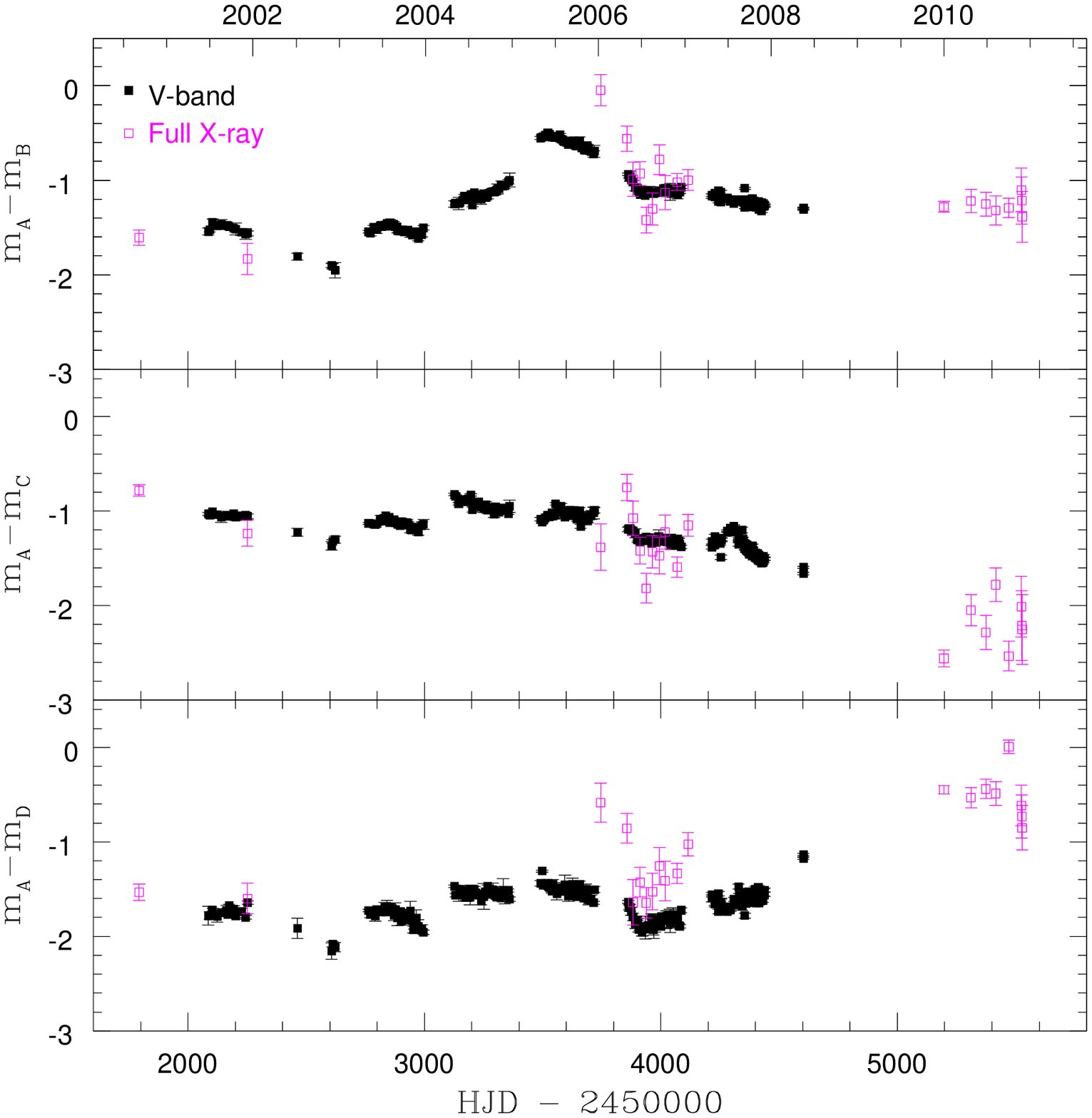}
\caption{\label{LC2237_VF} Optical (V-band, filled squares) and X-ray ($0.4-8.0$ keV, open squares) microlensing 
light curves for Q~2237$+$0305.}
\end{center}
\end{figure}

\begin{figure}
\begin{center}
\vspace{0.5 cm}
\includegraphics[scale=0.8]{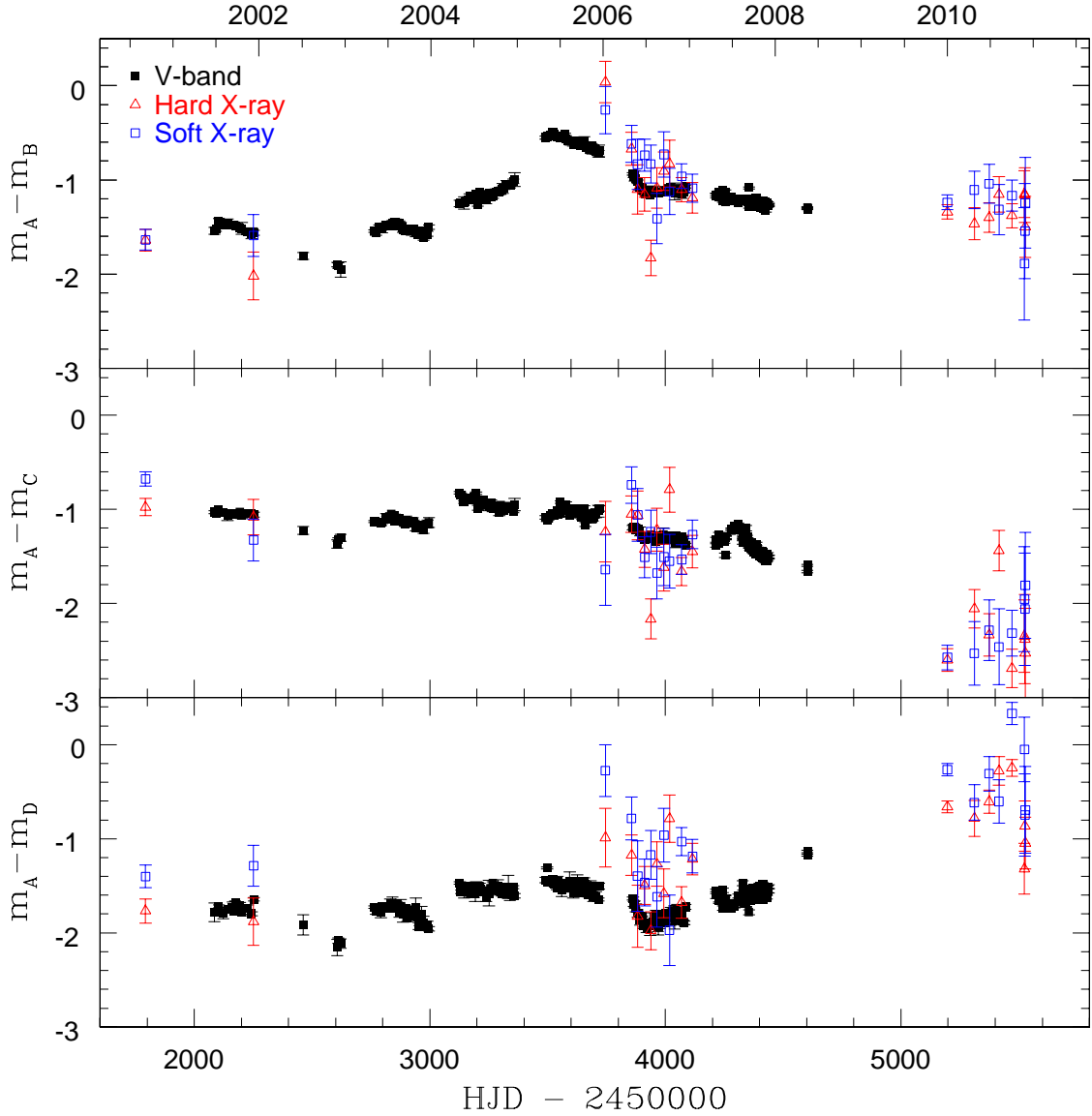}
\caption{\label{LC2237} Optical and X-ray microlensing light curves for Q~2237$+$0305. Filled squares represent 
V-band, and open squares and triangles soft ($0.4-1.3$ keV) and hard X-ray ($1.3-8.0$ keV) respectively.}
\end{center}
\end{figure}

\begin{figure}
\begin{center}
\vspace{0.5 cm}
\includegraphics[scale=0.8]{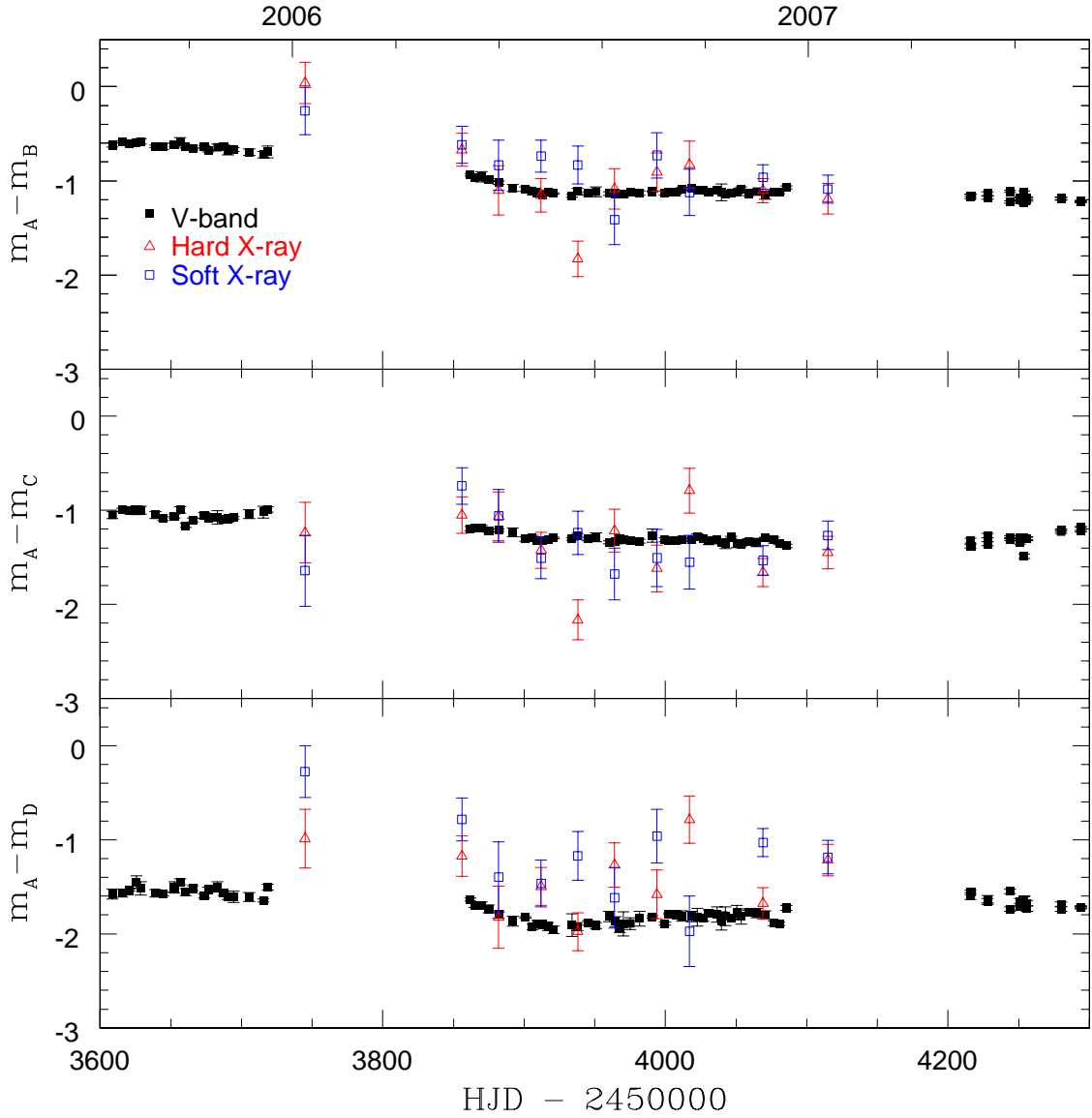}
\caption{\label{zoomC7} Closer view of the {\it Chandra} light curves during Cycle 7. The OGLE 
V-band data overlapping this period of time are also shown for comparison. Filled squares represent 
V-band, and open triangles and squares hard and soft X-ray bands respectively.}
\end{center}
\end{figure}

The optical data consist of the V-band observations from OGLE.
The lightcurve for our analysis goes from JD 2085 (June 2001) to JD 4602 (May 2008), and consists of 247 epochs. Details of the OGLE data 
reduction and photometric techniques can be found in \cite{Udalski}. We include the estimates of 
additional systematic errors in the photometry of 0.02, 0.03, 0.04 and 0.05 mag for images A, B, C and D respectively, 
following \cite{PK1, PK2}. Figures \ref{LC2237_VF} and \ref{LC2237} show the flux ratios for the B, C, and D images relative to A. 
In Fig. \ref{zoomC7} we have zoomed in 2006 (Cycle 7), where the X-rays show a significant variation over a short period of time, 
possibly due to a caustic crossing event.

\subsection{Magnification Patterns and Light curve fitting}

To fit the microlensing lightcurves we applied the Bayesian Monte Carlo Method of \cite{CK04}  
and \cite {PK1}, closely following the procedures in \cite{jeff2012f}. The basic idea 
is as follows. We built magnification patterns 
and convolved them with a source model to produce simulated light curves. We compared these 
light curves to the data and then use a Bayesian analysis to determine the parameters and their uncertainties.
When fitting multiple-wavelength data sets, we first fit the best 
sampled lightcurve (OGLE V-band in our case), and use the same trials to sequentially fit the other bands 
(hard and soft X-rays here) assuming a new source size for each band. We next describe these steps in greater detail.

For the models of the optical data we generated magnification patterns for each image of 4096$\times$4096 pixels, with an outer 
scale of $20\times20 \ R_E$, where for Q~2237$+$0305 the Einstein radius is $R_E=1.8\times10^{17}\langle M/M_{\odot}\rangle ^{1/2}$cm. 
This gives a resolution of $0.005 \ R_E/$pixel $= 9.0\times 10^{14} \langle M/M_{\odot}\rangle ^{1/2}$ cm/pixel. 
For comparison, \cite{morgan10} estimated a single epoch CIV BH mass of $10^{8.95}M_{\odot}$ with some caveats about 
the quality of the spectrum, and \cite{Roberto11} estimated masses of $10^{9.1}M_{\odot}$ and $10^{9.4}M_{\odot}$ for H$\alpha$ and H$\beta$ 
respectively, so the gravitational radius of the BH is roughly 
$r_g=GM_{BH}/c^2 \approx 1.5\times 10^{14} (M_{BH}/10^9 M_{\odot} )$ cm, and in this paper we eventually find an 
optical disk scale length of $ \sim 1.5\times 10^{16}$ cm. We used the same surface density, 
$\kappa$, and shear, $\gamma$, values as were used by \cite{PK1}, and fixed 
the mean microlensing mass to $\langle M\rangle=0.3M_{\odot}$, since this is roughly correct for older stellar populations, 
and any small variation in mass would only produce a constant shift $\propto \langle M \rangle ^{1/2}$ 
in the marginalized distribution of the projected size. Since for this system the surface density is 
dominated by the stars rather than by dark matter, we assumed that all the surface mass density is in stars, with  $\kappa_{\ast}/\kappa=1$.
In their earlier models of Q~2237$+$0305 varying these quantities \cite{PK1} found  $\langle M\rangle\sim 0.3M_{\odot}$, and 
\cite{CK04} found $\kappa_{\ast}/\kappa \sim 1$, consistent with these expectations.

The patterns change with time due to the motions of the Earth, the lens and its stars, and the source. For the Earth we 
combine the heliocentric CMB dipole and Earth's orbit projected onto the lens plane. We used the source and 
lens galaxy peculiar velocities estimated from \citet[][]{ana11b}. We assigned the stars a random velocity 
dispersion of 170 km s$^{-1}$ and velocity rotations of $55$ km s$^{-1}$ for images A and B, and of $20$ km s$^{-1}$ 
for images C and D based on \citet[][]{trott2010}.

Finally, the magnification pattern for each image and epoch is convolved with a simple thin disk 
model \citep{ss1973} without including an inner edge. We adopted a simple profile since microlensing 
is much more sensitive to the source size than to the shape of the brightness profile \citep{mortonson05, congdon07}, 
and we ignored the inner edge because it is generally small compared to the measured (optical) disk sizes. The optical disk scale 
length, $R_{\lambda}$, is defined by the radius at which $k_B T_{eff}(r) = hc/\lambda$, where $\lambda$ is the observed wavelength. 
The X-rays, however, are better characterized by the half-light radius, $R_{1/2}=2.44 R_{\lambda}$, and for simplicity we used the same 
brightness profile since there seems no reason to change it based on \cite{mortonson05} and \cite{congdon07}. While 
microlensing can constrain disk structure, it is generally achieved by measurements of size as a function of 
wavelength \citep[e.g.,][]{morgan12, jeff1104} rather than distinguishing between detailed structures at fixed wavelength. 

Since the OGLE V-band light curve is the best sampled, consisting of 247 epochs from JD 2085 to JD 4602, we 
fit it first. We traced $10^{5}$ random trajectories for each source structure, parametrizing the source models 
by choosing from projected areas covering $\mathrm{log}_{10}$(area/cm$^2$) = 29.4 to 34.6 in steps of $0.2$, at each of 10 inclinations, 
$i$, with $\mathrm{cos}\ i$ values going from 0.1 to 1.0 (face-on), and 18 disk major axis position angles from 
0$^\circ$ to 170$^\circ$ in steps of 10$^\circ$.
To account for differential dust extinction \citep[e.g.,][]{Falco1999,eigenbrod08b,Agol09}, 
undetected substructure \citep[e.g.,][]{mao, metcalf, CKND, Vegetti}, uncertainties in the macro models, and any other 
contamination from the lens or host galaxy, the time-constant relative magnifications of the images were allowed to vary 
with a Gaussian prior set to a 1.0 mag dispersion.

We saved all the parameters for trials with $\chi^2/N_{dof} \leq 5$. For each of these trial parameters 
we recompute the lightcurves for the next band, hard X-rays, but again varying the source size (reduced 
to a $\mathrm{log}_{10}$(area/cm$^2$) range of 28.8 to 34.2), inclination angle, and relative magnification offsets. 
Since the X-ray emission is more compact, we doubled the resolution of the patterns to $8192\times8192$ pixels 
while maintaining the outer scale of $20 R_E$ following the procedures developed by \cite{jeff2012f}. This 
gives a pixel scale of $2.4\times10^{14}$cm/pixel, which is small enough to resolve the average sizes found in the runs. 
We choose to fit the hard band next because its variability amplitude, especially during 2006, appears to be larger than observed 
in the soft band. This will reduce the computational time by reducing the number of solutions found at this stage, 
and therefore the number of initial conditions for the next band. In later runs of this band sequence we can be 
less ``liberal'' in keeping solutions. We keep hard X-ray fits with $\chi^2/N_{dof} \leq 4$ and then fit the soft 
X-ray band saving results with $\chi^2/N_{dof} \leq 3$. Dropped solutions would contribute to our Bayesian probability 
integrals as $\mathrm{exp}[-(\chi^2 -\chi^2_{min})/2]$, so the dropped cases should be exponentially 
unimportant to the final results.

\begin{figure}
\begin{center}
\vspace{0.5 cm}
\includegraphics[scale=0.8]{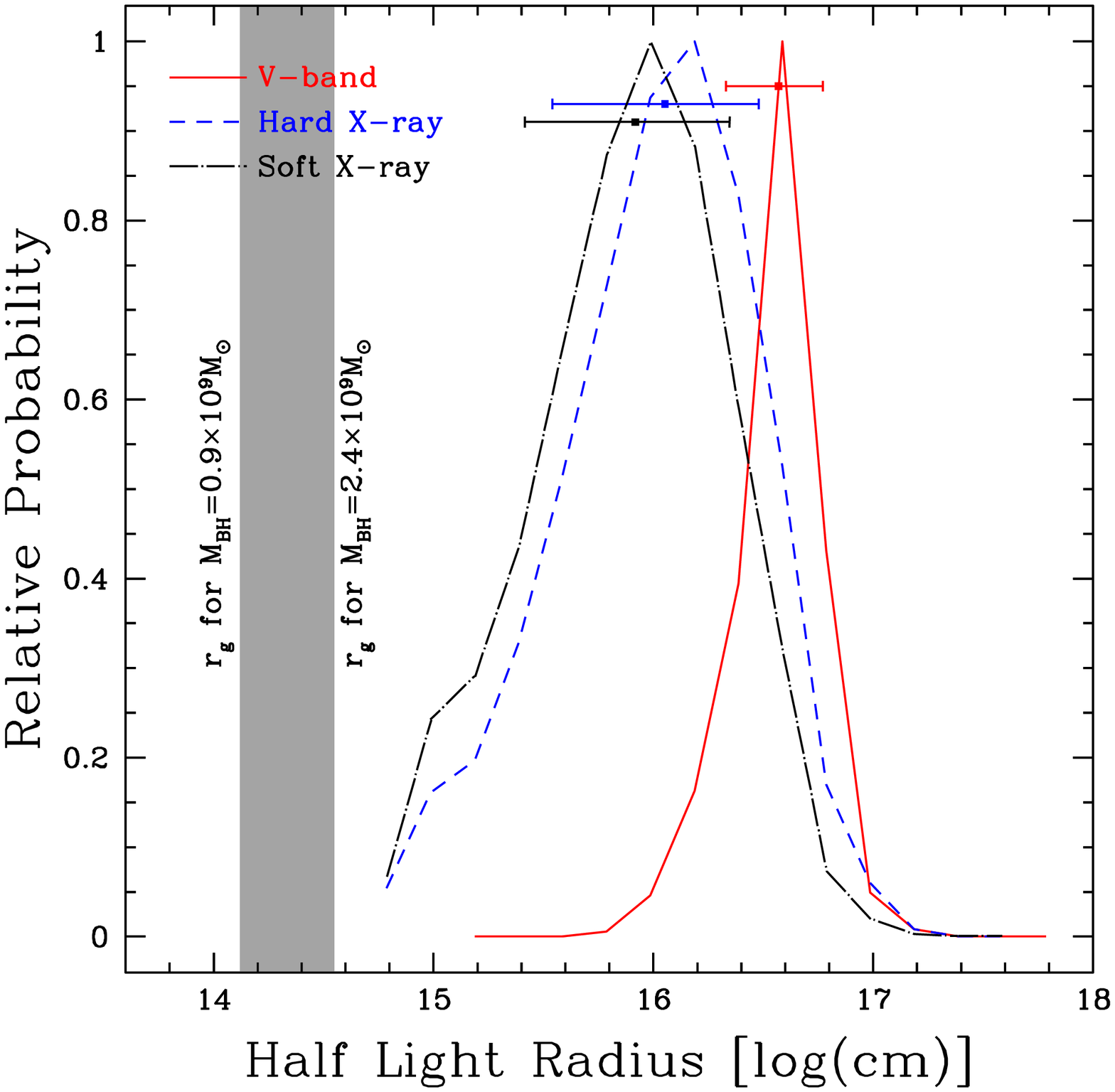}
\caption{\label{completeLC} Probability distributions for the half-light radii of the V-band (solid), 
soft X-ray (dash-dotted), and hard X-ray (dashed) emission. The gray area corresponds 
to predicted values of $r_g$, where the lower and upper limits are for $M_{\mathrm{BH}}=0.9\times 10^{9}M_{\odot}$ 
\citep[CIV,][]{morgan10} and $M_{\mathrm{BH}}=2.4\times 10^{9}M_{\odot}$ \citep[H$\beta$,][]{Roberto11} respectively.}
\end{center}
\end{figure}

\begin{figure}
\begin{center}
\vspace{0.5 cm}
\includegraphics[scale=0.8]{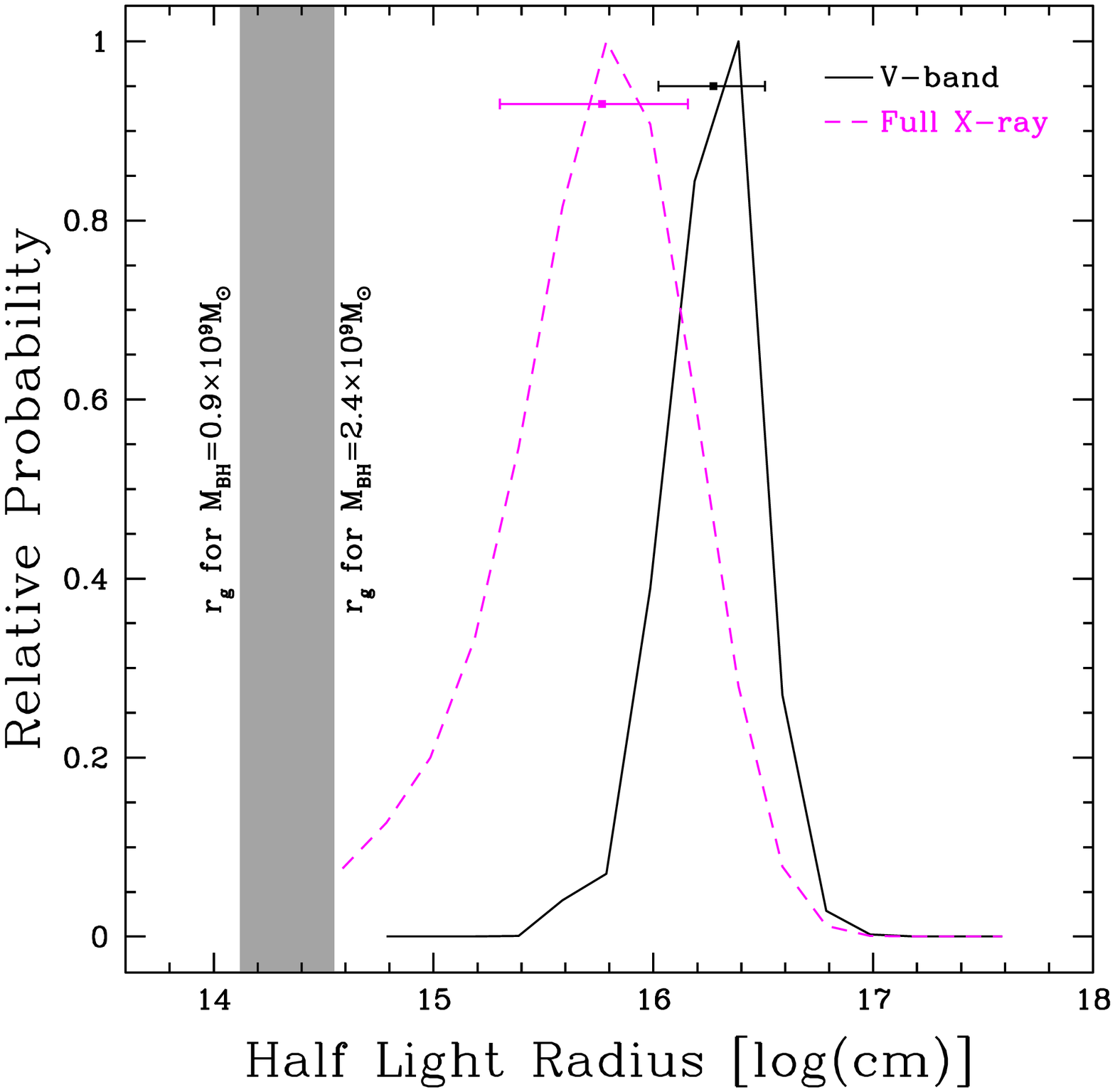}
\caption{\label{completeLCVF} Probability distributions for the half-light radii of the V-band (solid) and full 
X-ray (dashed) emission. The gray area corresponds to predicted values of $r_g$, where the lower and upper limits are 
for $M_{\mathrm{BH}}=0.9\times 10^{9}M_{\odot}$ \citep[CIV,][]{morgan10} and $M_{\mathrm{BH}}=2.4\times 10^{9}M_{\odot}$ \citep[H$\beta$,][]{Roberto11} 
respectively.}
\end{center}
\end{figure}

\begin{figure}
\begin{center}
\vspace{0.5 cm}
\includegraphics[scale=0.8]{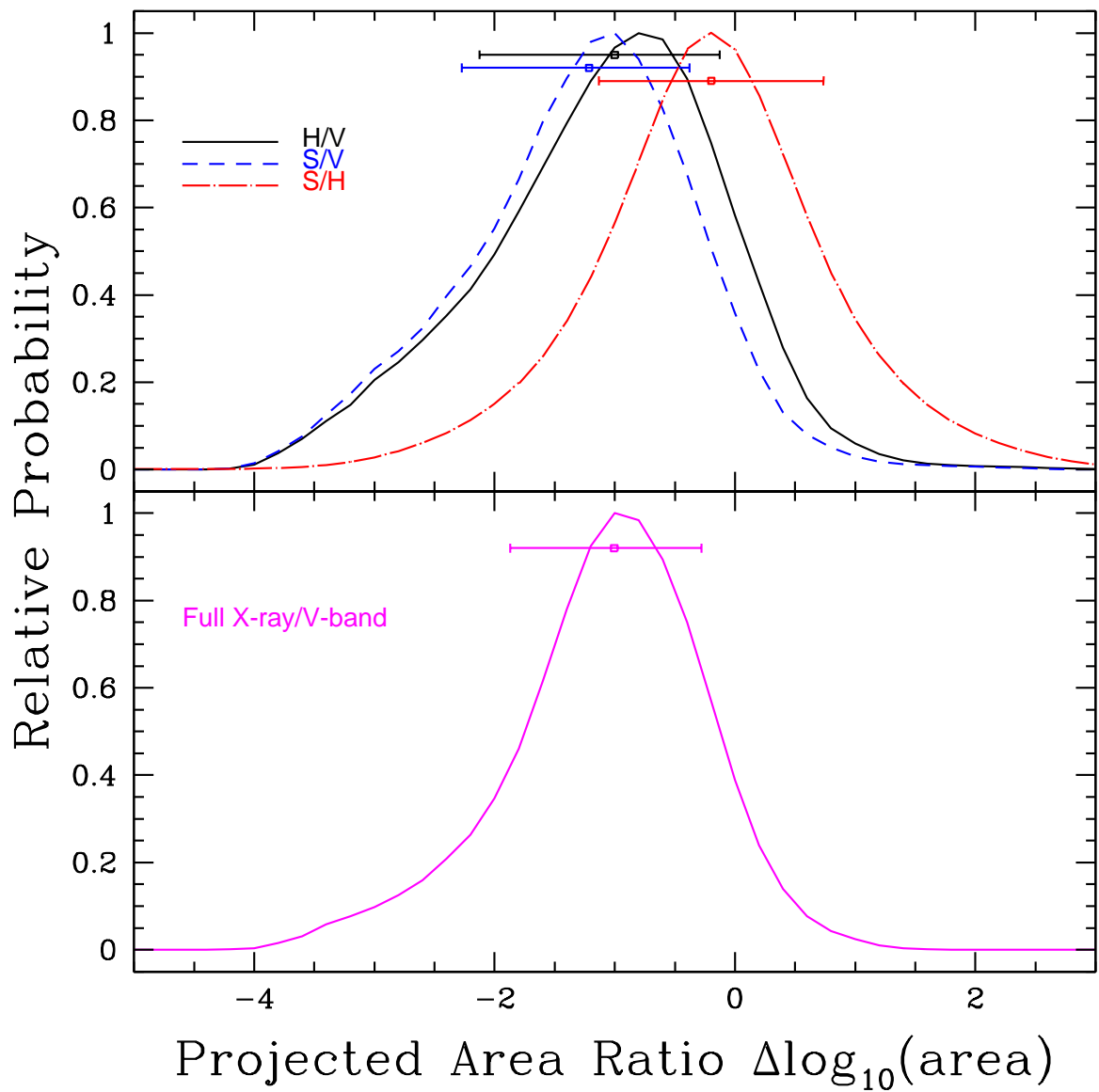}
\caption{\label{arearatio} Top panel: Probability distributions for the logarithm of the half-light radii hard X-ray area to optical area ratio (solid), 
along with similar distributions for soft X-ray to optical (dashed) and hard to soft X-ray (dash-dotted). Bottom panel: Similar 
distribution for the full X-ray area to optical area ratio.}
\end{center}
\end{figure}

Once the initial set of trials has been ``filtered'' through the three bands, we use a Bayesian analysis 
to determine the parameters and their uncertainties. Figure \ref{completeLC} shows the probability distribution 
for the scale radii obtained with a logarithmic prior on the sizes. We found that the X-ray emitting regions 
are significantly smaller than the optical, and found no significant difference between soft and hard bands, 
although the uncertainties are large. Between the different bands we found half-light radius ratios of 
$\mathrm{log}(R_{1/2,soft}/R_{V})=-0.65^{+0.47}_{-0.55}$, $\mathrm{log}(R_{1/2,hard}/R_{V})=-0.51^{+0.47}_{-0.57}$, 
and $\mathrm{log}(R_{1/2,soft}/R_{1/2,hard})=-0.14^{+0.60}_{-0.72}$. We also ran a sequence considering only the OGLE 
V-band and the full X-ray band, since the X-ray uncertainties are smaller when considering the complete energy range. We saved 
trials with full X-ray band fits with $\chi^2/N_{dof} \leq 5$, and we found 
$\mathrm{log}(R_{X,full}/R_{V})=-0.52^{+0.45}_{-0.54}$ (Fig. \ref{completeLCVF}). 
Thus the X-ray half-light radius is on scales of $\sim 10^{1.7\pm 0.5} \ r_g$, while the optical half-light radius is on scales of $10^{2.2\pm0.2}r_g$ 
for $M_{BH}=10^{9.1}M_{\odot}$. We also examined the distributions of the projected area ratios between the different wavelengths 
to see if the structure of the X-ray emitting region could be better defined (see Fig. \ref{arearatio}). 
These ratios have values of $\mathrm{log}(A_{X,full}/A_{V})=-1.01^{+0.73}_{-0.86}$, $\mathrm{log}(A_{X,soft}/A_{V})=-1.21^{+0.83}_{-1.07}$, 
$\mathrm{log}(A_{X,hard}/A_{V})=-1.01^{+0.86}_{-1.12}$, and $\mathrm{log}(A_{X,soft}/A_{X,hard})=-0.21^{+0.95}_{-0.91}$ that are consistent 
with the derived half-light radius ratios. 

Despite finding large numbers of good fits to the V-band data alone, when we included the X-ray data we had  difficulty obtaining 
large numbers of trials that were good fits to the data\footnote{Best $(\chi^2/N_{dof})_{hard} = 1.9$, and best $(\chi^2/N_{dof})_{soft} = 1.3$.}. 
Part of the problem is that it is probably 
exponentially harder to fit temporally longer light curves using the Monte Carlo approach \citep[see][]{PK1}. But another problem 
could be that we do not have fully aligned optical and X-ray data because we have no extension of the OGLE light curves to the most 
recent X-ray epochs. We tested this hypothesis by fitting the hard band alone. The number of good solutions we found was again small, 
although no significant change in the size estimate was observed when compared to the multi-band run. This suggests that the problem 
lies in the length of the X-ray lightcurves. One of the effects of finding small numbers of solutions is that the size at V-band 
is slightly shifted to larger values than found when fitting only the V-band data, because solutions 
with higher $\chi^2$ values dominate the weight after the multi-band run. Moreover, in the hard X-rays, many of the ``good'' solutions 
ignore the presence of the probable caustic crossing during 2006, whether only in the X-ray data or in the multi-band fits.

To examine these questions, we considered all the parameters from the V-band only fits (the complete OGLE light curve), 
but then separately analyzed different portions of the X-ray light curves. In one sequence we only fit the caustic 
crossing event occurring during 2006 (from JD 3745 to JD 4115), and in a second sequence we fit the points that 
characterize the long time scale changes between the {\it Chandra} cycles (JD 1793 and 2251 in 2000-2001, JD 3745 in 2006, 
and all the points in 2010, JD 5197 to 5527). Although analyzing them separately is less powerful in a statistical sense, we will not 
lose either the information encoded in the caustic crossing, or in the variability observed from 
cycle to cycle. Moreover higher quality solutions (i.e., with lower $\chi^2$) will be obtained for each multi-band run 
because we have decoupled the ``global'' statistics on long time scales from the ``local'' statistics of a particular 
caustic crossing event. We otherwise followed the same analysis procedures.

\begin{figure}
\begin{center}
\vspace{0.5 cm}
\includegraphics[scale=0.8]{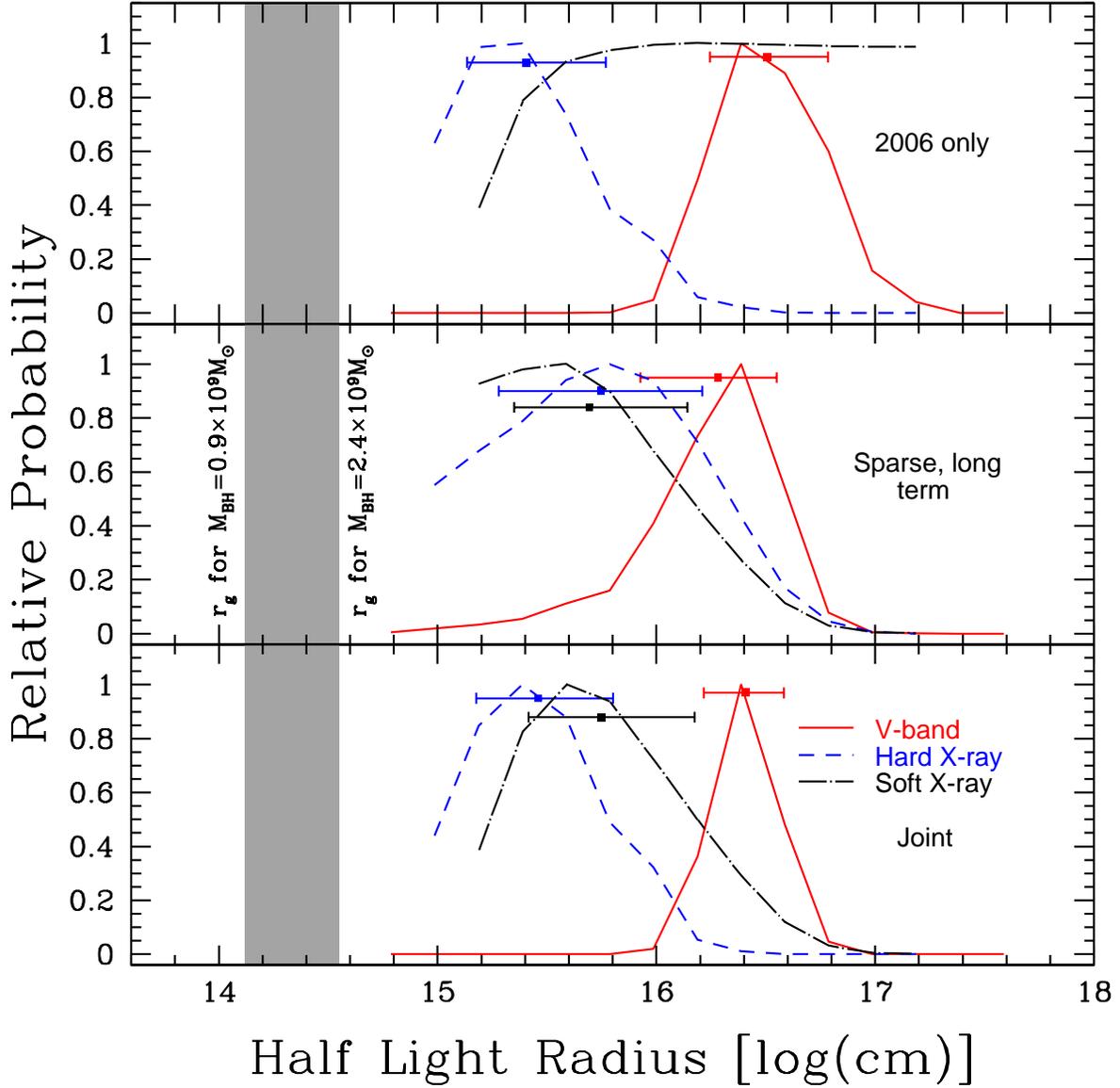}
\caption{\label{products} Probability distributions of the half-light radii for the V-band (solid), 
soft X-ray (dash-dotted), and hard X-ray (dashed) emission considering different sections of the X-ray light curves. 
The top panel shows the results corresponding to the analysis of the 2006 data, the middle panel to the sparse 
long term X-ray data, and the bottom panel shows the combined results. As in Fig. \ref{completeLC}, the gray 
area corresponds to predicted values of $r_g$.}
\end{center}
\end{figure}

The probability distributions for the scale radii for the independent sequences are shown 
in Figure \ref{products}. From the caustic crossing fitting results (2006 data, Figure \ref{products}, top panel) 
good constraints can be set on the hard X-ray emitting region\footnote {For the caustic crossing sequence the best $(\chi^2/N_{dof})_{hard} = 0.9$.} 
($\mathrm{log}(R_{1/2, hard}/\mathrm{cm})=15.41^{+0.35}_{-0.27}$), 
and as expected, due to it flatness, the soft band data does not provide any information about the size of the 
emission region in this energy range. The sparse long term X-ray data (Figure \ref{products}, 
middle panel), constrains both the soft and hard X-ray emitting 
regions\footnote {For the sparse long term sequence the best $(\chi^2/N_{dof})_{soft} = 1.02$, and the best $(\chi^2/N_{dof})_{hard} = 1.3$.}, 
although the uncertainties are large ($\mathrm{log}(R_{1/2, hard}/\mathrm{cm})=15.75^{+0.45}_{-0.48}$ 
and $\mathrm{log}(R_{1/2, soft}/\mathrm{cm})=15.70^{+0.43}_{-0.35}$). We can combine these results to improve 
our measurements by multiplying the probability distributions of the independent sequences (Fig. \ref{products}, bottom panel). 
For logarithmic size priors we find $\mathrm{log}(R_{1/2, hard}/\mathrm{cm})=15.46^{+0.34}_{-0.29}$, 
$\mathrm{log}(R_{1/2, soft}/\mathrm{cm})=15.76^{+0.41}_{-0.34}$, and $\mathrm{log}(R_{1/2, V}/\mathrm{cm})=16.41\pm0.18$. These 
results are consistent with the analysis of the full lightcurves, but we view them as more reliable given that we found 
a considerably higher number of good solutions for each sequence.

\section{Summary and Conclusions}\label{sec3}

Motivated by the recent detection of energy-dependent microlensing by \cite{chen2011, chen2012} in Q~2237$+$0305, 
we analyzed the OGLE V-band and \cite{chen2011, chen2012} {\it Chandra} data to study the structure of the optical and X-ray
emitting regions of this lensed quasar. We found that X-ray emission regions are significantly smaller than the optical, 
and considering two energy bands in the X-rays we estimated the half-light radii 
to be $\mathrm{log}(R_{1/2,hard}/\mathrm{cm})=15.46^{+0.34}_{-0.29}$, $\mathrm{log}(R_{X,soft}/\mathrm{cm})=15.76^{+0.41}_{-0.34}$, 
and $\mathrm{log}(R_{1/2,V}/\mathrm{cm})=16.41\pm0.18$. These results also suggest a structure for the X-ray 
emission region in which the hard component comes from a more compact region, although the statistical significance of 
this result is low. Since the geometry and extension of the disk depend upon the physics 
that triggers the X-ray emission, generally attributed to inverse Compton scattering of soft 
UV photons from the accretion disk by hot electrons in a corona surrounding the disk \citep[e.g.,][]{reynolds}, 
the smaller X-ray extent favors geometrical configurations like the light bending \citep{fabian05} or aborted jets 
models \citep{structureX}\, and  also agrees with the general relativistic MHD models of \cite{hirose} 
and  \cite{machida}.

\begin{figure}
\begin{center}
\vspace{0.5 cm}
\includegraphics[scale=0.8]{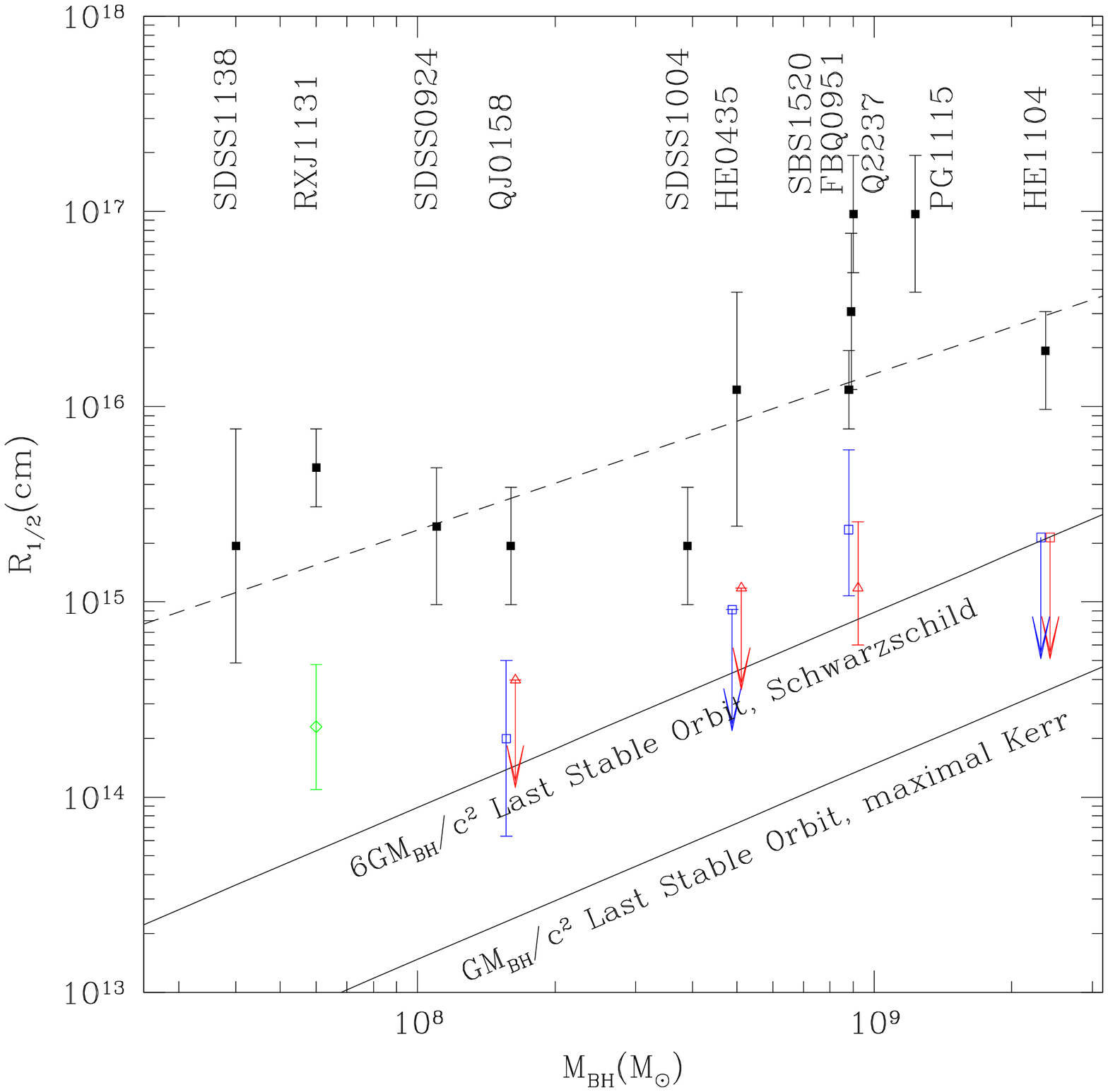}
\caption{\label{BHS} Emission region half-light radii as a function of black hole mass $M_{\mathrm{BH}}$ based in part on 
results from \cite{chartas09}, \cite{dai10}, \cite{morgan10,morgan12}, and \cite{jeff2012f, jeff1104}. 
Filled squares represent 
optical sizes, and open triangles and squares are for the hard and soft X-ray sizes, respectively. Small offsets between the 
soft and hard X-ray data were applied for clarity. Diamonds correspond to full band X-ray size estimates. Upper 
limits are indicated by arrows. The dashed line corresponds the \cite{morgan10} fit to the optical data.}
\end{center}
\end{figure}

We now have enough measurements to consider correlations between the X-ray size and $M_{BH}$. Fig. \ref{BHS} shows 
all the available optical and X-ray sizes following \cite{morgan10}. Here the optical 
results (filled squares) are from \cite{morgan06, morgan10, morgan12}, \cite{Fohlmeister}, \cite{dai10}, and \cite{jeff2012f, jeff1104}, 
the X-ray results (open symbols) are from \cite{dai10}, 
\cite{morgan12}, and \cite{jeff2012f, jeff1104}, and the BH mass estimates are from \cite{peng06}, \cite{morgan10}, \cite{greene10}, and \cite{Roberto11}.
Although X-ray size measurements exist for only a small number of systems, a first inspection suggests that the 
size of the X-ray emission region is roughly proportional to the mass of the central BH. Since the optical sizes appear to 
scale as the $R_{optical} \propto M_{BH}^{2/3}$ predicted by thin disk theory \citep{morgan10}, while the X-ray sizes appear to scale as 
$R_{X}\propto M_{BH}$, the differences between microlensing at X-ray and optical should be largest at low mass 
\citep[e.g., RXJ 1131$-$1231,][]{dai10} 
and modest at high mass (e.g., Q~2237$+$0305). However, the prevalence of upper limits on the X-ray sizes, as well as the 
large error bars, does not allow us to set strong constraints on the correlation. Because the X-ray emission is so compact, robust 
lower limits on sizes depend on having better sampled X-ray light curves than are typical of the existing data, although 
we are addressing this problem in a new set of Chandra observations.  With additional and better measurements, we can not only 
better constrain the correlation with $M_{BH}$ but also examine whether the spatial structure of the corona is correlated with the
X-ray spectral index, and explore the origin of the reflected X-ray components, particularly the Fe K$\alpha$ lines,
as we started to explore in \cite{chartas2012}.

\bigskip

\noindent Acknowledgments:
This research was supported by NASA/SAO grants
GO6-7093X, GO0-11121A/B/C, and GO1-12139A/B/C, 
and NSF grants AST-0708082 and AST-1009756. Further support for 
this work was provided by the National Aeronautics and Space Administration 
through Chandra Award Number 11121 issued by
the Chandra X-ray Observatory Center, which is operated by the Smithsonian 
Astrophysical Observatory for and on behalf of the National Aeronautics Space 
Administration under contract NAS8-03060.


\clearpage

\clearpage

\clearpage

\clearpage

\clearpage

\clearpage

\clearpage

\clearpage

\clearpage

\clearpage

\end{document}